\documentclass{article}
\usepackage{amsmath}
\usepackage{graphicx}
\usepackage{url}
\usepackage[footnotesize]{caption}

\begin{document}

\begin{center}
\large \textbf{STRINGY STABILITY OF DILATON BLACK HOLES IN 5-DIMENSIONAL ANTI--DE SITTER SPACE}
\end{center}
\begin{center}
\footnotesize{Yen Chin Ong}\\
\footnotesize{Department of Mathematics,\\ National University of Singapore,\\
Block S17 (SOC1), \\
10, Lower Kent Ridge Road, Singapore 119076.\\
E-mail: yenchin@nus.edu.sg\footnote{The author is now at Graduate Institute of Astrophysics and Leung Center for Cosmology and Particle Astrophysics, National Taiwan University at Taipei, and can be contacted at d99244003@ntu.edu.tw.}\\}
\end{center}

\begin{abstract}
Flat electrical charged black holes in 5-dimensional Anti--de Sitter space have been applied to the study of the phase diagram of quark matter via AdS/CFT correspondence. In such application it is argued that since the temperature of the quark gluon plasma is bounded away from zero, the dual black hole cannot be arbitrarily cold, but becomes unstable due to stringy instability once it reaches sufficiently low temperature. We study the stringy stability of flat dilaton black holes with dilaton coupling $\alpha=1$ in asymptotically Anti--de Sitter space and show that unlike the purely electrically charged black hole, these dilaton black holes do not suffer from stringy instability. 
\end{abstract}

\section{Holography of Flat Black Holes}

AdS/CFT correspondence has been employed to study the phase transitions of quark matter, where the temperature of the quark gluon plasma corresponds to electrical charges on a flat black hole in the 5-dimensional AdS bulk. Adding electrical charges to cool down the black hole eventually causes the black hole to become unstable \cite{Brett1,Brett2}. This instability, first discovered by Seiberg and Witten \cite{SW}, is due to the presence of brane in the bulk that modifies the geometry of the black hole (Euclidean) spacetime at large distances.
For a friendly introductory account of the Seiberg--Witten instability, see Ref. \cite{Kleban}. For simplicity, all electrical charges will be taken as positive in the subsequent discussion.\\

The \textit{Seiberg--Witten action} is given by 
\[
S_{\text{SW}} = \Theta\left(\text{Brane Area}\right) - \nu\left(\text{Volume Enclosed by Brane}\right)
\]
where $\Theta$ is the tension of the brane and $\nu$ relates to the charge enclosed by the brane. Instability arises if $S$ becomes negative since brane nucleation leads to large branes having energy which is unbounded from below as they approach the boundary. Equivalently, stability requires that the scalar curvature at conformal infinity should remain non-negative. This is clearly the case for AdS space itself, with positively curved conformal infinity. The most dangerous case is when the charge on the brane is saturated, called the BPS case. In 5 dimension, this means that $\nu_{\text{BPS}}=4\Theta/L$ where $L$ is the curvature scale of the AdS space. For the case of flat black hole, adding charges up to about 96\% of the extremal value effectively makes the Seiberg--Witten action becomes negative at large $r$, and destroys the stability of the black hole \cite{Brett1}. Consequently on the field theory side of the story, it means that quark gluon plasma cannot be arbitrarily cold, which is what we should expect. \\

In low energy limit of string theory, scalar field called \textit{dilaton}, denoted by $\phi$, can couple to the Maxwell field, giving the action
\[
S=\int d^4x \sqrt{-g} \left(R-2(\nabla \phi)^2 - e^{-2\phi}F^2\right).
\]
We will assume that the dilaton decays and vanishes at infinity. Note that because of the coupling between dilaton field and Maxwell field, the dilaton is not an independent ``hair'' of the black hole. \\

Recently, dilaton black holes have also been explored for its holography and applications in AdS/CFT correspondence \cite{Rocha, Goldstein, Chiangmei}. Notably Gubser and Rocha \cite{Rocha} argued that dilaton black hole in AdS$_5$ or a relative of it with similar behavior might be dual to Fermi liquid. Unlike quark gluon plasma, Fermi liquid can attain zero temperature. Therefore the dual black hole should be allowed to reach extremal charge without subjected to the Seiberg--Witten instability. In the following sections, we will first recall the properties of dilaton black holes, followed by computation of its Seiberg--Witten action to study its stability.

\section{Spherically Symmetric Dilaton Black Holes in Asymptotically Flat Spacetime}

We first recall 4-dimensional dilaton black holes in asymptotically flat case. For the sake of comparison we shall recall the Reissner--Nordstr\"om solution:
\[
g({\text{RN}}) = -\left(1-\frac{2M}{r}+\frac{Q^2}{r^2}\right)dt^2 + \left(1-\frac{2M}{r}+\frac{Q^2}{r^2}\right)^{-1}dr^2 + r^2d\Omega^2. 
\]
We assume $Q>0$ for simplicity.\\

For $0 < Q < M$, we have two horizons
\[
r_{\pm} = M \pm \sqrt{M^2 - Q^2}
\]
where $r=r_+$ is the event horizon and $r=r_-$ is the inner horizon. The corresponding black hole solution in the presence of dilaton field in low energy limit of string theory is remarkably simple. Known as the \textit{Garfinkle--Horowitz--Strominger} or GHS black hole \cite{GHS, Horowitz}, its metric is:
\[
g(\text{GHS}) = -\left(1-\frac{2M}{r}\right)dt^2 + \left(1-\frac{2M}{r}\right)^{-1}dr^2 + r\left(r-\frac{Q^2}{M}\right)d\Omega^2.
\]
As mentioned before, the dilaton is coupled to the electric field and hence is not an independent parameter of the black hole. The precise relation between the dilaton and the electric charge $Q$ is given by
\[
e^{-2\phi} = 1-\frac{Q^2}{Mr}.
\]
Note the absence of dependence on electrical charge in the $g_{tt}$ and $g_{rr}$ terms. The $r$-$t$ plane is thus similar to the Schwarzschild black hole, but the spherical horizon is smaller in area for any nonzero electrical charge.\\

Interestingly the GHS black hole behaves differently compared to the Reissner--Nordstr\"om black hole when electrical charge is increased. In the latter case, the event horizon moves \textit{inward} while the Cauchy horizon moves outward, finally the two horizon coincide when extremality is reached. For the GHS black hole however, the event horizon stays \textit{fixed} at $r_+ = 2M$ and it has \textit{no inner horizon}. The effect of decreasing electrical charge on the GHS black hole is to decrease its area, which goes to zero at $Q^2 = rM$. This gives the extremal limit: the \textit{event horizon becomes singular} at $Q^2 = 2M^2$, i.e. $Q=\sqrt{2}M$, unlike the extremal limit of Reissner--Nordstr\"om black hole that satisfies $Q=M$. One can understand that the dilaton black hole has larger charge over mass ratio in the extremal limit because the scalar field contributes an extra attractive force, and so for any fixed $M$, we need a larger $Q$ to balance it.\\

In general, we can introduce a free parameter $\alpha \geq 0$, that governs the strength of coupling between the dilaton field and the Maxwell field. This yields the \textit{Garfinkle--Maeda} or GM black hole solution \cite{ChengZhouLiu}:
\begin{flalign*}
g(\text{GM}) = &-\left(1-\frac{r_+}{r}\right)\left(1-\frac{r_-}{r}\right)^{\frac{1-\alpha^2}{1+\alpha^2}}dt^2 + \left(1-\frac{r_+}{r}\right)^{-1}\left(1-\frac{r_-}{r}\right)^{\frac{\alpha^2-1}{\alpha^2+1}}dr^2 \\ & +r^2\left(1-\frac{r_-}{r}\right)^{\frac{2\alpha^2}{1+\alpha^2}}d\Omega^2
\end{flalign*}
where
\[
e^{-2\phi} = e^{-2\phi_0}\left(1-\frac{r_-}{r}\right)^{\frac{2\alpha}{1+\alpha^2}}, ~~ F=\frac{Q}{r^2}dt \wedge dr
\]
with the asymptotic value of the dilaton field $\phi_0$ taken to be zero in the following discussion. The horizons are at 
\[
r_{\pm} = \frac{1+\alpha^2}{1\pm \alpha^2}\left[M \pm \sqrt{M^2 - (1-\alpha^2)Q^2}\right], \alpha \neq 1 ~~\text{for}~ r_-.
\]
When $\alpha =1$ (the coupling strength that appears in the low energy string action), the GM solution reduces to the GHS solution, and there ceases to be an inner horizon, while $\alpha = 0$ case reduces to the Reissner--Nordstr\"om solution. 

\section{Topological Dilaton Black Holes}

Gao and Zhang \cite{GaoZhang} generalized the GM solution to include dilatonic topological black hole in asymptotically AdS spacetime in $n$-dimension, 
\[
ds^2 = -U(r)dt^2 + W(r)dr^2 +[f(r)]^2 d\Omega_{k,n-2}^2
\]
where $k=-1,0,+1$ and
\[
U(r) = \left[k-\left(\dfrac{r_+}{r}\right)^{n-3}\right]\left[1-\left(\dfrac{r_-}{r}\right)^{n-3}\right]^{1-\gamma(n-3)} - \dfrac{1}{3}\Lambda r^2 \left[1-\left(\dfrac{r_-}{r}\right)^{n-3}\right]^\gamma \]
\begin{flalign*}
W(r) = &\left\{\left[k-\left(\dfrac{r_+}{r}\right)^{n-3}\right]\left[1-\left(\dfrac{r_-}{r}\right)^{n-3}\right]^{1-\gamma(n-3)} - \dfrac{1}{3}\Lambda r^2 \left[1-\left(\dfrac{r_-}{r}\right)^{n-3}\right]^\gamma \right\}^{-1}\\& \times \left[1-\left(\dfrac{r_-}{r}\right)^{n-3}\right]^{-\gamma(n-4)}\\
\end{flalign*}
with
\[
\left[f(r)\right]^2 = r^2 \left[1-\left(\dfrac{r_-}{r}\right)^{n-3}\right]^{\gamma}; ~~~~~~ \gamma = \dfrac{2\alpha^2}{(n-3)(n-3+\alpha^2)}.
\]
\\

Note that in this notation $\Lambda$ is the effective cosmological constant $|\Lambda| = 3/L^2$ where $L$ is the curvature scale of de Sitter or Anti--de Sitter space, independent of dimensionality. We also note that for $n \geq 5$, $\alpha \neq 0$, we have $U(r)W(r)\neq 1$ in general. This is not surprising since the presence of scalar field contributes to the stress energy tensor and thus affects the geometry of spacetime leading to $g_{tt}g_{rr} \neq -1$ in general \cite{Ted}. \\

The mass of the black hole, and the \textit{charge parameter} $q$, are, with $L=1$ for simplicity, \\
\[
M= \frac{\Gamma_{n-2}}{16\pi} (n-2)\left[r_+^{n-3} + k\left(\frac{n-3-\alpha^2}{n-3+\alpha^2}\right)r_-^{n-2}\right]\\
\]
and
\[
q^2 = \frac{(n-2)(n-3)^2}{2(n-3+\alpha^2)}r_+^{n-3}r_-^{n-3}.
\] 
respectively. The charge parameter $q$ is directly proportional to the black hole electrical charge $Q$ \cite{Hendi}.\\

Let us consider $k=0$, $L=1$, $n=5$, $\alpha = 0 = \gamma$ which should reduce to the case of flat charged black hole in Ref. \cite{Brett1,Brett2}. Using the above formula, we compute that
\[
M = \frac{8\pi^3K^3}{16\pi}3r_+^2 = \frac{3}{2}\pi^2 K^3 r_+^2
\]
and 
\begin{flalign*}
U(r) & = \left(-\left(\frac{r_+}{r}\right)^2\right)\left(1-\left(\frac{r_-}{r}\right)^2\right)-\frac{1}{3}\Lambda r^2 \\
&= -\frac{r_+^2}{r^2} + \frac{r_+^2r_-^2}{r^4} + \frac{r^2}{L^2} = W(r)^{-1}.
\end{flalign*}\\

If we compare this with the explicit form of metric for flat electrically charged black hole, 
\[
U(r) = \frac{r^2}{L^2} - \frac{2M}{3\pi^2K^3r^2} + \frac{Q^2}{48\pi^5K^6 r^4}
\]
we see that 
\[
r_+^2 = \frac{2M}{3\pi^2 K^3}, ~~ r_+^2r_-^2 = \frac{Q^2}{48\pi^5K^6}.
\]
The event horizon is the solution of $U(r)=0$ which is \textit{not} the square root of $\frac{2M}{3\pi^2K^3}$. Thus in the notation of Gao and Zhang, $r_+$ and $r_-$ are merely parameters that relate to the horizons instead of the horizons themselves. The authors in Ref.\cite{Hendi} for example, use the symbols $c$ and $b$ in place of $r_+$ and $r_-$ and refer to them as ``integration constants''. 

\section{Seiberg--Witten Action for Flat AdS Dilaton Black Holes}

Consider the 5-dimensional flat dilaton black hole in AdS with $L=1$. As above,
\[
r_+^2 = \frac{2M}{3\pi^2K^3}.
\] 
Since $Q^2$ is proportional to $q^2$, we have
\[
Q^2 \equiv Q^2(\alpha) = Q^2(\alpha =0)\frac{n-3}{n-3+\alpha^2} = Q^2(\alpha =0)\frac{2}{2+\alpha^2}.
\]
That is,
\[
Q^2 = \frac{2}{2+\alpha^2}(48\pi^5K^6r_+^2r_-^2).
\]
I.e.
\[
r_-^2 = \frac{Q^2(2+\alpha^2)}{96\pi^5K^6r_+^2}= \frac{Q^2(2+\alpha^2)}{96\pi^5K^6\left(\frac{2M}{3\pi^2K^3}\right)}.
\]
\\

We study the comparatively easy case of $\alpha = 1$ in which we have
\[
Q^2 = 32\pi^5K^6r_+^2r_-^2, ~~ r_-^2 = \frac{3Q^2}{64M\pi^3K^3},
\]
\[
\gamma = \frac{2(1)}{2(2+1)}=\frac{1}{3}, ~~ f(r)^3 = r^3\left[1-\left(\frac{r_-}{r}\right)^2\right]^{\frac{1}{2}}.
\]
Thus the Euclidean metric satisfies
\begin{flalign*}
g_{\tau\tau} &= \left[-\left(\frac{r_+}{r}\right)^2\right]\left[1-\left(\frac{r_-}{r}\right)^2\right]^{1-\frac{2}{3}} + r^2\left[1-\left(\frac{r_-}{r}\right)^2\right]^{\frac{1}{3}}\\
& = \left[-\left(\frac{r_+}{r}\right)^2 + r^2\right]\left[1-\left(\frac{r_-}{r}\right)^2\right]^{\frac{1}{3}}
\end{flalign*}
and
\begin{flalign*}
g_{rr} =  \left\{\left[-\left(\frac{r_+}{r}\right)^2 + r^2\right]\left[1-\left(\frac{r_-}{r}\right)^2\right]^{\frac{1}{3}}\right\}^{-1}\left(1-\left(\frac{r_-}{r}\right)^2\right)^{-\frac{1}{3}}.
\end{flalign*}
with the horizon at
\[
r_{\text{eh}} =(r_+^2)^{\frac{1}{4}}=\left(\frac{2M}{3\pi^2K^3}\right)^{\frac{1}{4}}
\]
which is fixed independent of the electrical charge, just like its asymptotically flat GHS cousin. \\

For any fixed dilaton coupling $\alpha$, varying the electrical charge means equivalently, varying the parameter $r_-$, via the relationship
\[
r_-^2 = \frac{3Q^2}{64M\pi^3K^3}.
\]
At extremal limit, the horizon becomes singular with $r_\text{eh}=r_+^{\frac{1}{2}} = r_-$, i.e.
\[
\left(\frac{2M}{3\pi^2K^3}\right)^{\frac{1}{4}} = \left[\frac{3Q^2}{64M\pi^3 K^3}\right]^{\frac{1}{2}}.
\]
Therefore the extremal charge is 
\[
Q_E=\frac{8\times 2^{\frac{1}{4}}}{3^{\frac{3}{4}}} \pi M^{\frac{3}{4}} K^{\frac{3}{4}} \approx 13.11~M^{\frac{3}{4}} K^{\frac{3}{4}} .
\]
Again this is greater than $Q_E \approx 9.96 (KM)^{\frac{3}{4}}$ for flat AdS Reissner--Nordstr\"om black hole \cite{Brett1}, a similar behavior as its asymptotically flat counterpart.\\

The Seiberg--Witten action $S_{\text{SW}}$ takes the form:
\[
\alpha\left\{r^3\left[1-\frac{3Q^2}{64\pi^3K^3Mr^2}\right]^{\frac{2}{3}}\left[r^2-\frac{2M}{3\pi^2K^3r^2}\right]^{\frac{1}{2}} - 4\int_{r_{\text{eh}}}^r dr' (r')^3 \left[1-\frac{3Q^2}{64\pi^3K^3M(r')^2}\right]^{\frac{1}{3}}\right\}
\]
where $\alpha=2\pi \Theta PLA_k$. Here $2\pi P$ is the period imposed on the imaginary time after Wick-rotation and $A_k$ the area of the event horizon. 
The action vanishes at the horizon. See details of the technique in Ref.\cite{Brett1}.\\

Taking typical values of the parameters we can obtain a plot of the action as function of $r$ as in Fig.~1. 
\begin{center}
\begin{figure}
\includegraphics[width=4.5 in]{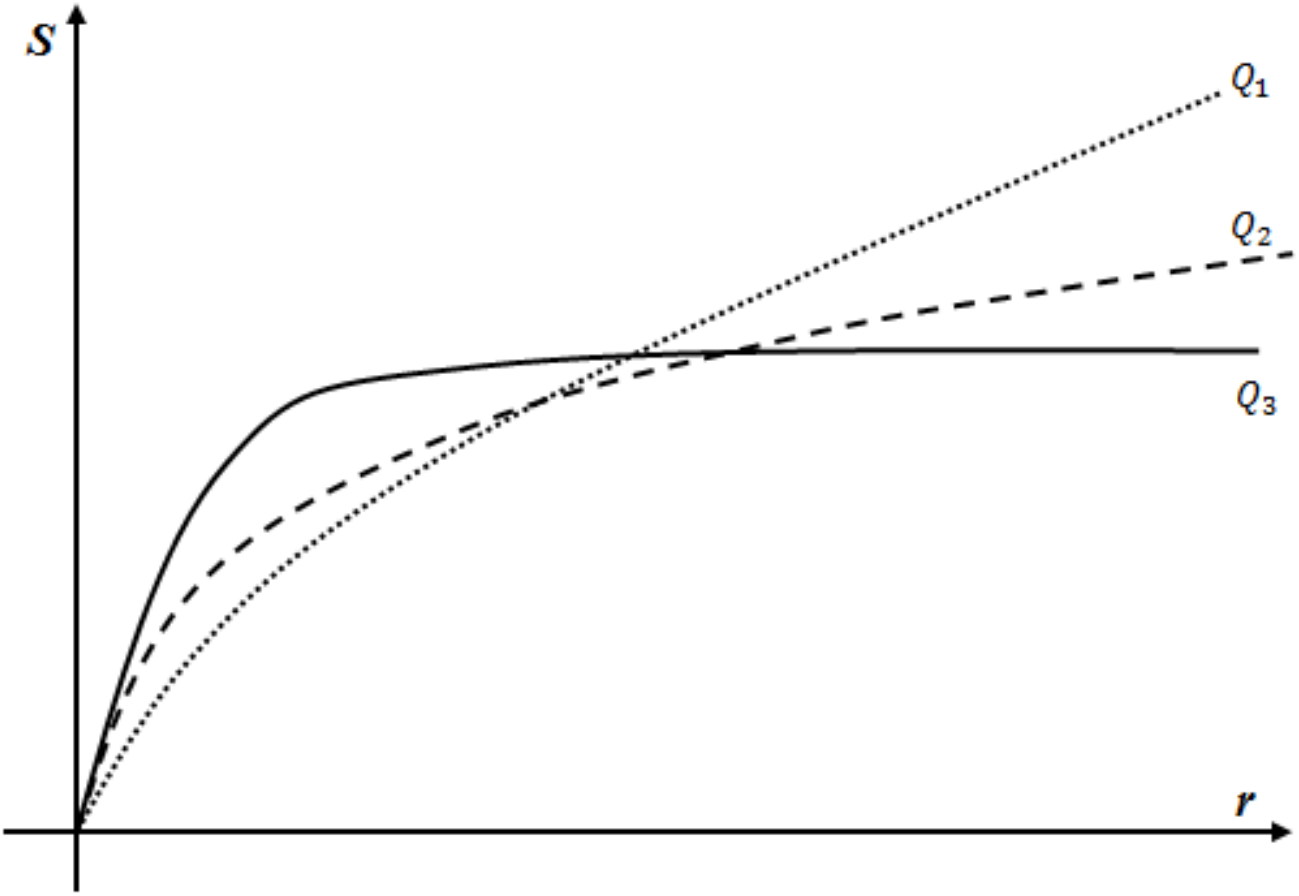}
\caption{The Seiberg--Witten action with typical values of parameter, where $Q_1 > Q_2 > Q_3 \approx 0$. The curve for $Q=0$ would correspond to flat uncharged black hole case where the action asymptotes to a constant positive value. $S=0$ axis corresponds to the horizon.}
\end{figure}
\end{center}

We note that for $Q=0$ the action reduces to that of uncharged flat AdS black hole (the dilaton, being a secondary hair coupled to the Maxwell field, also vanishes when electrical charge is zero), which asymptotes to a positive value \cite{Brett1}. Unlike flat AdS Reissner--Nordstr\"om black hole with action increases to a maximum before plunging to negative, the action of flat AdS dilaton black hole is always positive. In particular, $\displaystyle \lim_{r\to\infty} S(r,Q_E) = +\infty$ where $Q_E$ denotes the extremal charge. For any fixed charge $Q$, increasing the charge makes the action starts out with smaller value than the one with charge $Q$, but subsequently takes over at some finite value of $r$. The value of $r$ in which this take over occurs decreases with increasing charge. Thus the presence of dilaton stabilizes the black hole (at least in this special case with $\alpha = 1$) against non-perturbative instability in the Seiberg--Witten sense. \\

\section{Conclusion}

If AdS/CFT correspondence were to make sense between flat dilaton black hole and condensed matter system which can attain zero temperature, then the black hole should not become unstable in the Seiberg--Witten sense when electrical charges are increased. We have shown that this is indeed so at least for the case of dilaton black hole with coupling $\alpha=1$. More works are needed to study the stringy stability of dilaton black holes with general coupling strength, especially for those as yet unknown range of values that are of application interest in AdS/CMT. The work of Hendi and Sheykhi \cite{Hendi} suggests that dilaton black holes possess unstable phase for large value of $\alpha$, at least thermodynamically speaking. The conjecture of Gubser and Mitra \cite{GubserMitra} would suggest that such black holes are also dynamically unstable, although more works are still required to settle this stability issue definitively. 

\section*{Acknowledgments}

The author wishes to express his deep gratitude to his thesis supervisor Professor Brett McInnes for useful discussions and advises without which this work would not have been possible.

\end{document}